# Decameter Type III-Like Bursts

V.N. Mel'nik (1), A.A. Konovalenko (1), H.O. Rucker (2), B.P. Rutkevych (1), V.V. Dorovskyy (1), E.P. Abranin (1), A.I. Brazhenko (2), A.A. Stanislavskii (1), A. Lecacheux (3)

1- Institute of Radio Astronomy, Ukrainian Academy of Sciences, Kharkov, Ukraine
2 - Space Research Institute, Austrian Academy of Sciences, Graz, Austria
3 - Gravimetrical Observatory, Poltava, Ukraine
4 - Departement de Radioastronomie, Observatoire de Paris, Paris, France

**Abstract** We report the first observations of Type III-like bursts at frequencies 10 – 30 MHz. More than 1000 such bursts during 2002 – 2004 have been analyzed. The frequency drift of these bursts is several times that of decameter Type III bursts. A typical duration of the Type III-like bursts is 1 – 2 s. These bursts are mainly observed when the source active region is located within a few days from the central meridian. The drift rate of the Type III-like bursts can take a large value by considering the velocity of Type III electrons and the group velocity of generated electromagnetic waves.

## 1. Introduction

The fast Type III bursts (Type III-like bursts) were observed for the first time in 1959 (Young *et al.*, 1961) at frequencies of 500 – 950 MHz. The frequency drift rates of these bursts exceeded 2000 MHz $s^{-1}$ and sometimes were virtually infinite. Bursts with reverse drift (from low to high frequencies) were also observed, with a relative occurrence rate of about 20% of the observed Type III-like bursts. This rate is greater than that of inverted U-bursts, so that these bursts cannot be attributed to the descending branch of the inverted U-bursts. The duration of high-frequency Type III-like bursts varies from 0.3 to 2 s and faster bursts tend to have shorter durations. The radio fluxes of these bursts were in the range from 5 to $10^4$ s.f.u. (where 1 s.f.u. $= 10^{-22}$ W $\mathrm{Hz}^{-1}$ $\mathrm{m}^{-2}$).

Another report (Kundu *et al.*, 1961) pointed out that Type III-like bursts were detected in the frequency range of 400 – 800 MHz but sometimes their frequencies extended to as low as 200 MHz and to as high as 950 MHz. These bursts typically occurred in groups of three to ten bursts. Kundu *et al.* (1961) also paid attention to the fact that some bursts had drift rates of different signs at low and high frequencies.

A more detailed analysis of the properties of Type III-like bursts observed at frequencies of 310 – 340 MHz was carried out by Elgaroy and Rosenkilde (1974) and Elgaroy (1980). By considering a few hundred Type III-like bursts he concluded that 50%, 17%, and 33% of bursts had forward, reverse, and infinite drifts, respectively. The mean duration of Type III-like bursts was 0.26 s or just a quarter of the duration of usual Type III bursts (1.1 s) in the same frequency range. Following Kundu *et al.* (1961), Elgaroy (1980) noted that these bursts occurred in groups and that the mean time interval between them was 1.1 s. He also considered the relationship between the drift rate and duration of Type III-like bursts and the position of active regions associated with them. It turned out that, as the associated active region moved off the central meridian, the burst duration increased and the fraction of Type III-like bursts with infinite and reverse drift rates decreased. Elgaroy (1980) concluded that these properties could be explained by the propagation effects of electromagnetic waves.

At lower frequencies, 40 – 50 MHz, Gopala Rao (1965) investigated the polarization of Type III bursts with different durations. He found that the shortest bursts with 1 – 2 s duration (which could be related to Type III-like bursts although their drift rates were not defined in the paper) had high polarization, up to 70%.

Zaitsev and Levin (1984) made an attempt to find Type III-like bursts in the decameter range by analyzing observational data of a storm on 8 – 9 June 1977. They concluded that, when bursts were detected, their duration and drift rate were similar to those for usual Type III bursts.

In the papers of Young *et al.* (1961) and Elgaroy (1980) the explanation of the Type III-like burst phenomenon was based on the idea that fast electron streams generating these bursts traveled in the solar corona with large density gradients. In their interpretation, the large drift rates of Type III-like bursts can easily be explained. If the electrons moved toward the Sun, bursts with reverse drifts could be observed.

Recently, Ledenev (2000) proposed that the positive frequency drift could be explained by a decrease in the signal group delay as the emission source moved in the direction of decreasing density. He also drew attention to the fact that to calculate mutual delay of signals arriving from different levels in the corona it was necessary to consider a sufficiently high velocity of Type III electrons.

According to Zaitsev and Levin's (1984) model the same electron streams were the sources for both the Type III-like bursts and the standard Type III bursts, but the former were connected with the fastest electrons in the stream and the latter with the main body of this stream. In this case, the ratio in drift rates of the fast and the standard Type III bursts should be approximately a factor of two. In addition, in this model the Type III-like bursts should appear mainly in the low corona at high frequencies.

In the present paper we report the first results of observations of Type III-like bursts in the decameter range. Their properties are similar to those for high-frequency Type III-like bursts. We note a strong dependence of the occurrence rate of Type III-like bursts with regard to the active region position on the solar disk. This fact suggests a decisive role of the propagation effects of radio emission in the phenomenon of decameter Type III-like bursts.

## 2. Observations

The Type III-like bursts studied in this paper were observed with the radio telescope UTR-2 in the years 2002 – 2004. Three sections of the radio telescope UTR-2 with a total area of 30 000 $m^2$ were used, providing a beam of $1^\circ \times 13^\circ$. In 2002 the observation was carried out with the digital spectrograph DSP (Kleewein, Rosolen, and Lecacheux, 1997), whose frequency and time resolutions were 12 kHz and 20 ms, 50 ms, and 100 ms, respectively. The frequency bandwidth of the DSP was 12 MHz from 18 to 30 MHz. Additionally, the 60-channel spectrometer with time resolution selectable up to 10 ms and a frequency resolution fixed at 300 kHz in the frequency band of 10 – 30 MHz was also used in the 2002 – 2004 observations.

In the decameter range, 10 – 30 MHz, the usual Type III bursts have frequency drift rates of 2 – 4 MHz $s^{-1}$ and durations of 4 – 10 s (Abranin *et al.*, 1980, 1990). But at the same time, during Type III burst storms we sometimes observed fast Type III bursts with drift rates considerably exceeding 2 – 4 MHz $s^{-1}$ (Mel'nik *et al.*, 2006). As a rule their durations were smaller than those for usual Type III bursts. Following Elgaroy (1980) we also call them Type III-like bursts. In this paper more than 1000 Type III-like bursts, which were observed during five storm events on 13 – 17 and 24 – 30 July 2002, 16 – 19 August 2002, 1 – 6 July 2003, and 17 – 21 June 2004, have been considered (Table 1). The total time of analyzed observations is more than 130 hours.

Practically all Type III-like bursts have negative drifts from high to low frequencies (Figure 1a). Their time profile is symmetrical (Figure 1b), which is different from the asymmetrical time profile of usual Type III bursts with a steep rise and a slow decrease of the flux density. There are also Type III-like bursts with fine frequency structure similar to usual Type IIIb bursts. An example of such Type IIIb-like bursts is shown in Figure 2. These bursts are not observed at high frequencies; they are typical for the decameter range. Our analysis shows that the majority of Type III-like bursts are observed when the active region associated with them is near the central meridian. The occurrence rate of Type III-like bursts is presented for the storm on 24 – 30 July 2002 in Figure 3a. We see that on 28 July, when the active region was slightly east from the central meridian, the occurrence rate of bursts was 10 times higher than for the previous and the following days.

The same situation was found for the 1 – 6 July 2003 storm (Figure 4), when an active region crossed the central meridian on 4 – 5 July 2003. In the case of the 13 – 17 July 2002 storm the occurrence rate reached a maximum on 16 July (Table 1), when an active region intersected the central meridian. In the storm on 16 – 20 August 2002, two active regions passed through the central meridian with a two-day interval (Figure 5). Exactly at those days there are two maxima in the occurrence rate of Type III-like bursts. For the storm of 17 – 21 June 2004 with two active regions the situation is analogous. So we can conclude that the phenomenon of Type III-like bursts seems to exist because of the propagation effects of electromagnetic radiation through the corona, confirming Elgaroy's (1980) results. We also note a correlation between the occurrence rate of Type III-like bursts and the size of an active region – increasing size implies higher occurrence rate.

**Table 1.** Number and occurrence rate of Type III-like bursts observed during five storm events of Type III bursts in 2002 – 2004.

| Date of observation | Number of bursts per | Duration of observation | Number of bursts |
|---|---|---|---|
| 13.07.2002 | 0.021 | 5:36 | 7 |
| 14.07.2002 | 0.027 | 4:57 | 8 |
| 15.07.2002 | 0.035 | 4:48 | 10 |
| 16.07.2002 | 0.057 | 3:47 | 13 |
| 17.07.2002 | 0.003 | 3:40 | 1 |
| 26.07.2002 | 0.073 | 5:28 | 24 |
| 27.07.2002 | 0.088 | 6:17 | 33 |
| 28.07.2002 | 0.457 | 6:14 | 171 |
| 29.07.2002 | 0.018 | 5:37 | 6 |
| 30.07.2002 | 0.058 | 2:52 | 10 |
| 16.08.2002 | 0.038 | 7:00 | 16 |
| 17.08.2002 | 0.117 | 3:26 | 24 |
| 18.08.2002 | 0.062 | 7:30 | 28 |
| 19.08.2002 | 0.077 | 1:18 | 6 |
| 01.07.2003 | 0.004 | 7:30 | 2 |
| 03.07.2003 | 0.091 | 2:56 | 16 |
| 04.07.2003 | 0.403 | 6:20 | 153 |
| 05.07.2003 | 0.277 | 4:39 | 77 |
| 06.07.2003 | 0.028 | 6:31 | 11 |
| 17.06.2004 | 0.066 | 5:04 | 20 |
| 18.06.2004 | 0.195 | 6:05 | 71 |
| 19.06.2004 | 0.453 | 5:44 | 156 |
| 20.06.2004 | 0.333 | 6:03 | 121 |
| 21.06.2004 | 0.409 | 5:20 | 131 |
| 22.06.2004 | 0.187 | 5:32 | 62 |

According to our analysis (Figure 6a), the Type III-like bursts have frequency drifts of mainly 5 – 10 MHz $s^{-1}$, but there is a considerable number of them with drifts up to 20 – 30 MHz $s^{-1}$. The maximum frequency drift rate observed was 42 MHz $s^{-1}$ for the burst on 18 August 2002 (Figure 7). Thus the drift rates of Type III-like bursts can exceed those for usual

Type III bursts by a factor of 10 or more. The drift rate of one selected Type III-like bursts is practically constant in the whole frequency band from 10 to 30 MHz, in contrast to the general case of Type III bursts in which the drift rate increases with frequency (Suzuki and Dulk, 1985). Type III-like bursts have durations of approximately 1 – 2 s (Figure 6b), which is one-quarter to one-fifth those for the usual Type III bursts. Therefore the situation is similar to the case of high-frequency bursts (Young *et al.*, 1961; Elgaroy, 1980). The flux of decameter Type III-like bursts exceeds 10 s.f.u. and sometimes 100 s.f.u. (Figure 6c). We could not find any connection between the fluxes and the drift rates. Corresponding histograms for other storms are similar to those shown in Figure 6. The only Type III-like burst with positive drift rate was observed on 13 July 2002 (Figure 8). This is in contrast to high-frequency Type III-like busts, in which positive drift rates are common. The drift rate, duration, and flux for this particular burst (Figure 8) are 27 MHz $s^{-1}$, 1 s, and 50 s.f.u., respectively.

## 3. Discussion

Until recently Type III-like bursts had been considered to appear mainly in the frequency range from 200 to 950 MHz. Our observations now show that these bursts occur also at decameter wavelengths. The presence of such fast bursts in the decameter range is a challenge to the theory of this phenomenon. At high frequencies Type III-like bursts are usually connected with inhomogeneities in the coronal plasma. This looks quite reasonable for the low solar corona. But in the case of decameter Type III-like bursts when the drift rate is constant from 30 to 10 MHz, the size of the corresponding inhomogeneity should be comparable to the solar radius. Moreover, its density gradient should be constant and several times higher than the density gradient of the surrounding solar corona, because the Type III burst emission observed at the same time starts exactly there. In our opinion the model proposed by Zaitsev and Levin (1984) also cannot explain these decameter bursts because their model predicts that the Type III-like burst would have a drift rate approximately twice that of the usual Type III bursts. According to the observations there are a number of bursts with considerably higher drift rates. If one supposes that their model works, then it would be necessary to admit that Type III-like bursts with drift rates of 20 – 30 MHz s$^{-1}$ are generated with superluminal velocities. By using the standard equation (Zheleznyakov, 1964)

$$(df\ /\ dt)_s \approx df\ /\ dn \cdot dn\ /\ dr \cdot v_s \tag{1}$$

which connects the drift rate d*f*/d*t* and the source (electron) velocity $v_S$, we again derive superluminal velocities for Type III-like electrons for the mean solar corona. Both the models of Young *et al.* (1961) and Elgaroy (1980) and the model of Zaitsev and Levin (1984) also cannot explain the strong relation of observed burst number on the location of an active region. This last fact indicates the particular role of the propagation effect of electromagnetic waves.

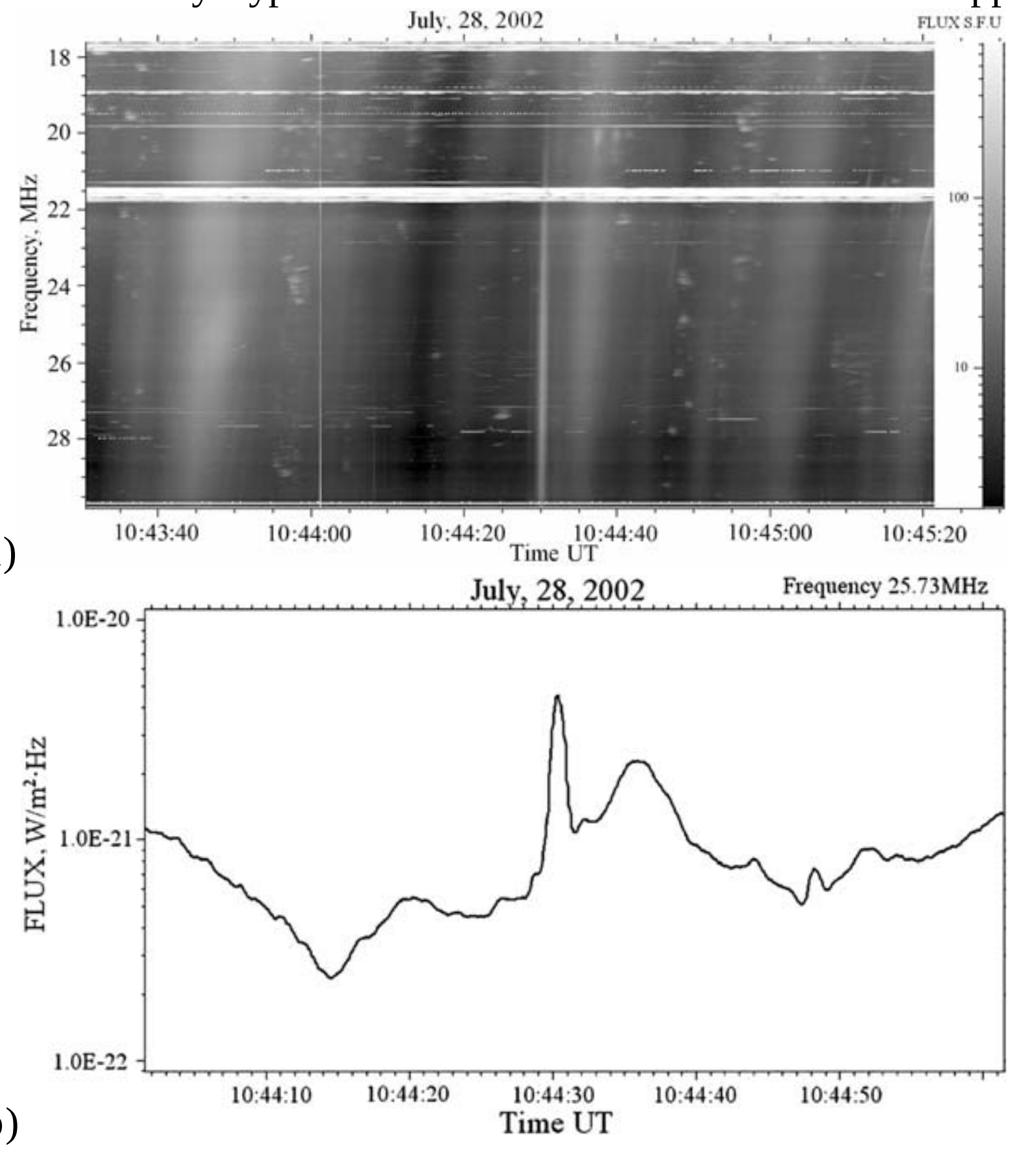

a)

b)

**Figure 1.** (a) Type III-like burst (10:44:30 UT) against a background of a Type III burst storm; (b) its time profile.

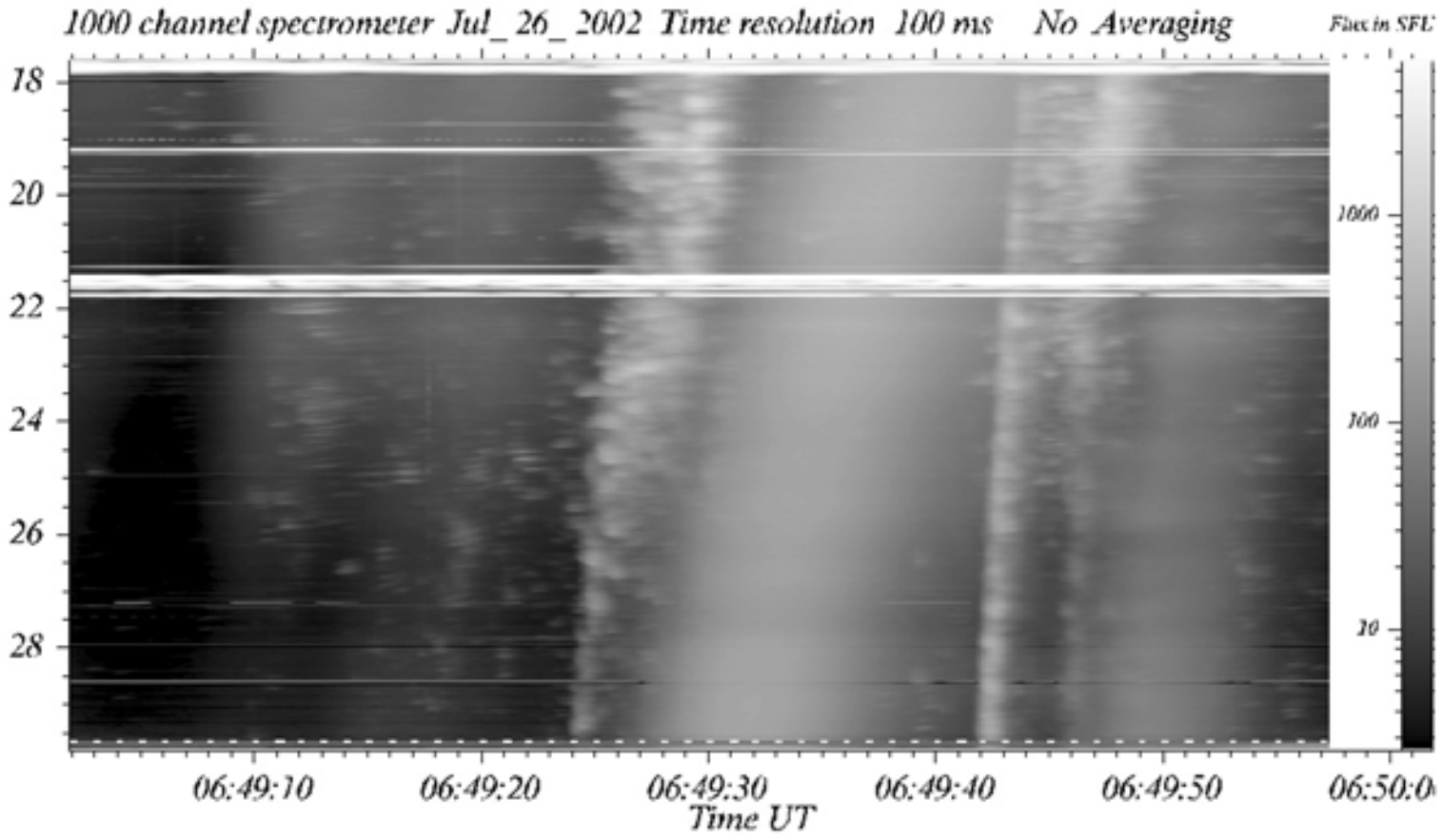

**Figure 2.** A Type IIIb-like burst (06:49:42 UT) similar to the standard Type IIIb bursts.

Following Ledenev's (2000) idea, we take into account the velocity of Type III electrons when deriving the equation for the frequency drift and show that this provides a natural explanation for the large drift rates found in Type III-like bursts. Let us consider the drift rate of the radio emissions generated by an electron beam at two close points. The electrons that propagate with velocity $v_S$ in the direction of an angle $\alpha$ to the line of sight will successively generate electromagnetic waves at points 1 and 2 (Figure 9). The time difference between the arrivals of these waves to the observer is

$$\Delta t = t_2 - t_1 = \frac{r_{12}}{v_s} + \int_{R_1}^{R_2} \frac{dr}{v_{gr}} \approx \frac{r_{12}}{v_s \cdot v_{gr}}(v_{gr} - v_s \cos\alpha) \tag{2}$$

where $r_{12}$ is the distance between points 1 and 2, and $v_{gr}$ is the group velocity of the electromagnetic wave generated by the electron stream. Equation (2) is obtained in the approximation of small angle $\alpha$. Then for the frequency drift rate we derive the following equation:

$$\frac{df}{dt} = \frac{f_2 - f_1}{t_2 - t_1} \approx \frac{df}{dn} \cdot \frac{dn}{dr} \cdot \frac{v_s \cdot v_{gr}}{(v_{gr} - v_s \cos\alpha)} \qquad (3)$$

a)

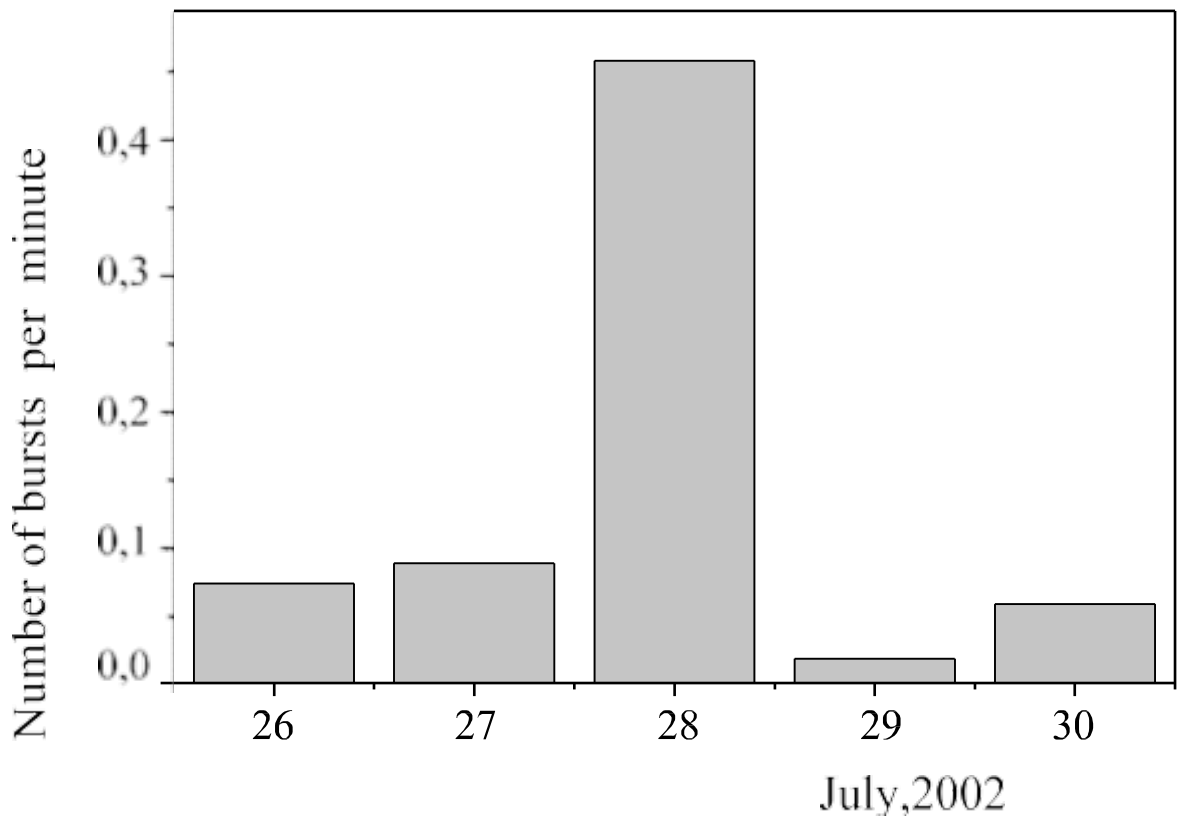

b)

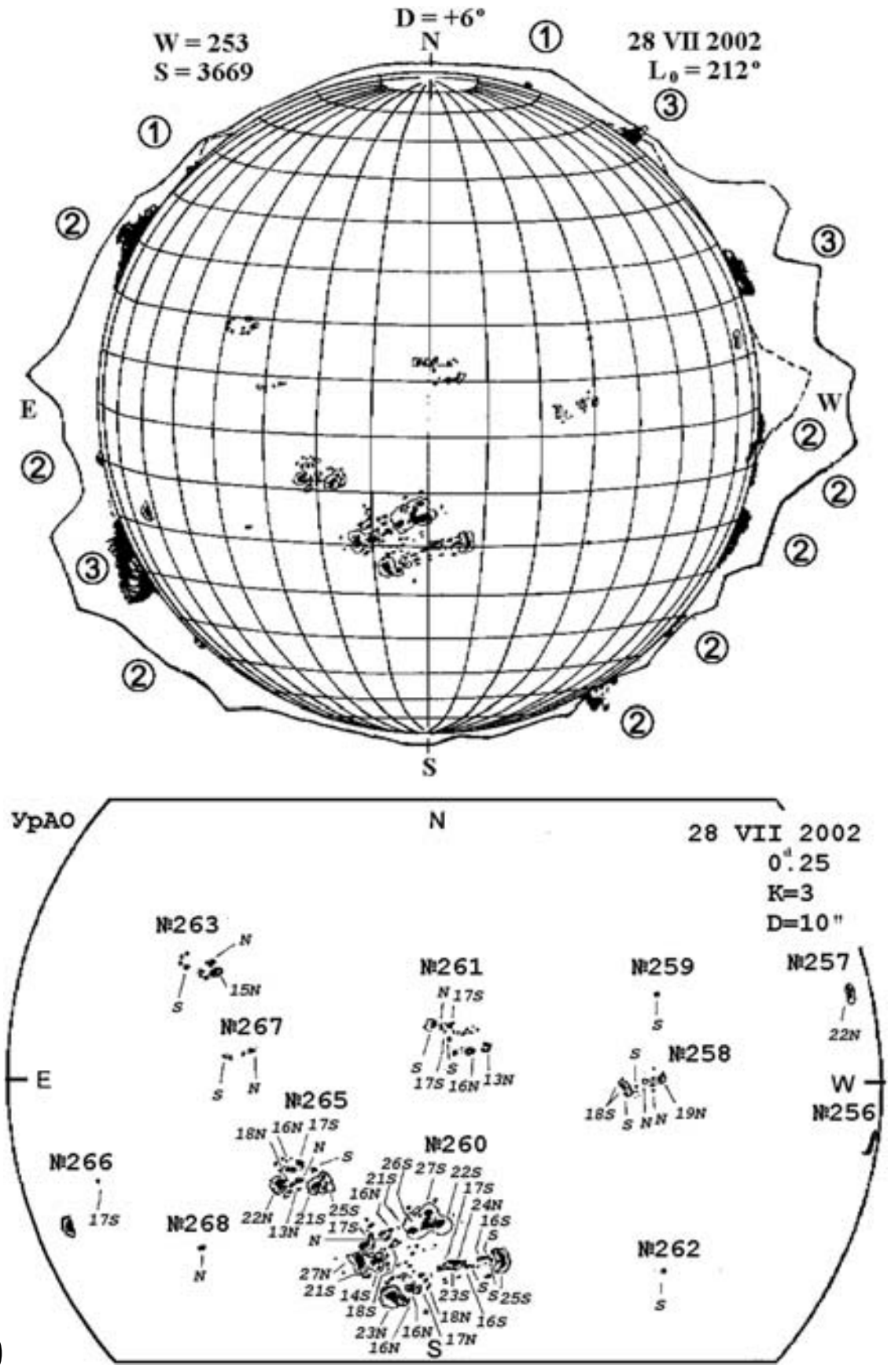

**Figure 3.** (a) The occurrence rate of Type III-like bursts for the storm event of 24 – 30 July 2002. (b) The location of the active region on the solar disk on 28 July 2002 (http://www.gao.spb.ru/).

In the case when $v_{gr} \approx c$ and $v_s \ll c$, Equation (3) turns into Equation (1). For further consideration let us ignore the density dependence in the model and introduce the value $\gamma$ defined by

$$\gamma = (\frac{df}{dt})_{gr} / (\frac{df}{dt})_s = \frac{v_{gr}}{(v_{gr} - v_s \cos\alpha)} \quad (4)$$

Here $\gamma$ depends on the group velocity of the electromagnetic wave and on the electron velocity but not on the coronal density. We see from Equation (4) that when the electron velocity is close to the group velocity then the drift rate can be large and it can even change sign.

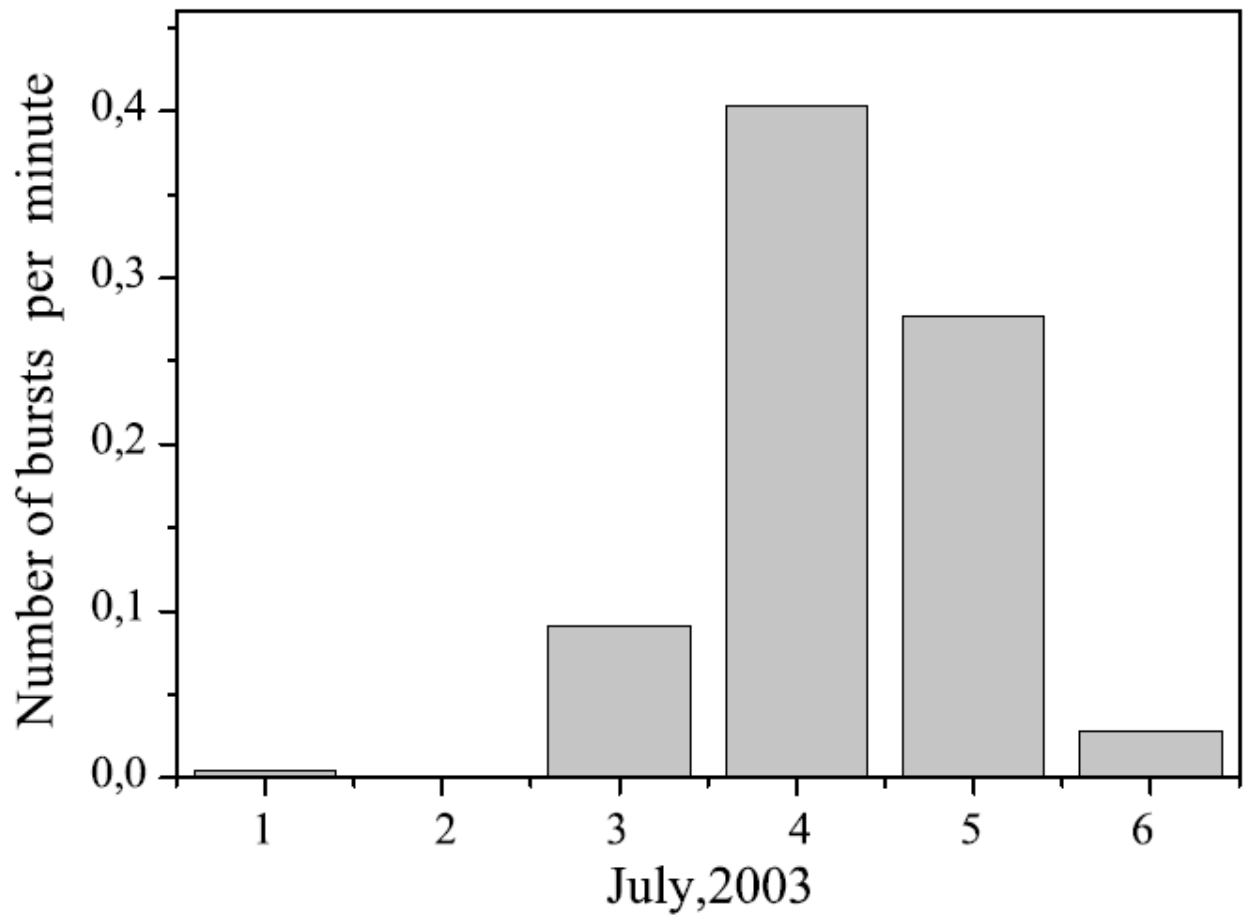

**Figure 4.** Occurrence rate of Type III-like bursts for 1 – 6 July 2003.

Let us consider this condition for the cases of fundamental and harmonic emissions in detail, supposing that the electrons with velocity $v_s$ generate Langmuir waves with wave number $k_l = \omega_{pe}/v_s$.

In the case of the generation of the fundamental radio emission (process $l + i \rightarrow t + i$, in which the Langmuir wave $l$ is scattered on the ion $i$ and is transformed into the electromagnetic wave $t$), the frequency of the electromagnetic wave $\omega_t$ is equal to the frequency of the Langmuir wave $\omega_l$. Therefore from the equation

$$\omega_t = \sqrt{\omega_{pe}^2 + k_t^2 c^2} = \omega_l = \sqrt{\omega_{pe}^2 + 3k^2 v_{Te}^2} \quad (5)$$

we derive the wave number of the electromagnetic wave as

$$k_t = \sqrt{3}\frac{v_{Te}}{c} k_l = \sqrt{3}\frac{v_{Te}}{c}\frac{\omega_{pe}}{v_s} \quad (6)$$

and its group velocity $v_{gr} = d\omega_t/dk_t = k_t c^2/\omega_t$ [using Equation (5)] as

$$v_{gr} = \frac{\sqrt{3} v_{Te} c}{v_s} \quad (7)$$

Taking into account Equations (4) and (7) we have

$$\gamma = \frac{\sqrt{3}v_{Te}c}{\sqrt{3}v_{Te}c - v_s^2 \cos\alpha} \tag{8}$$

The critical velocity

$$v_s{*} = \sqrt{\frac{\sqrt{3}v_{Te}c}{\cos\alpha}} = \frac{5\cdot 10^9}{\sqrt{\cos\alpha}}cms^{-1} \tag{9}$$

is the velocity at which $\gamma$ becomes infinite.

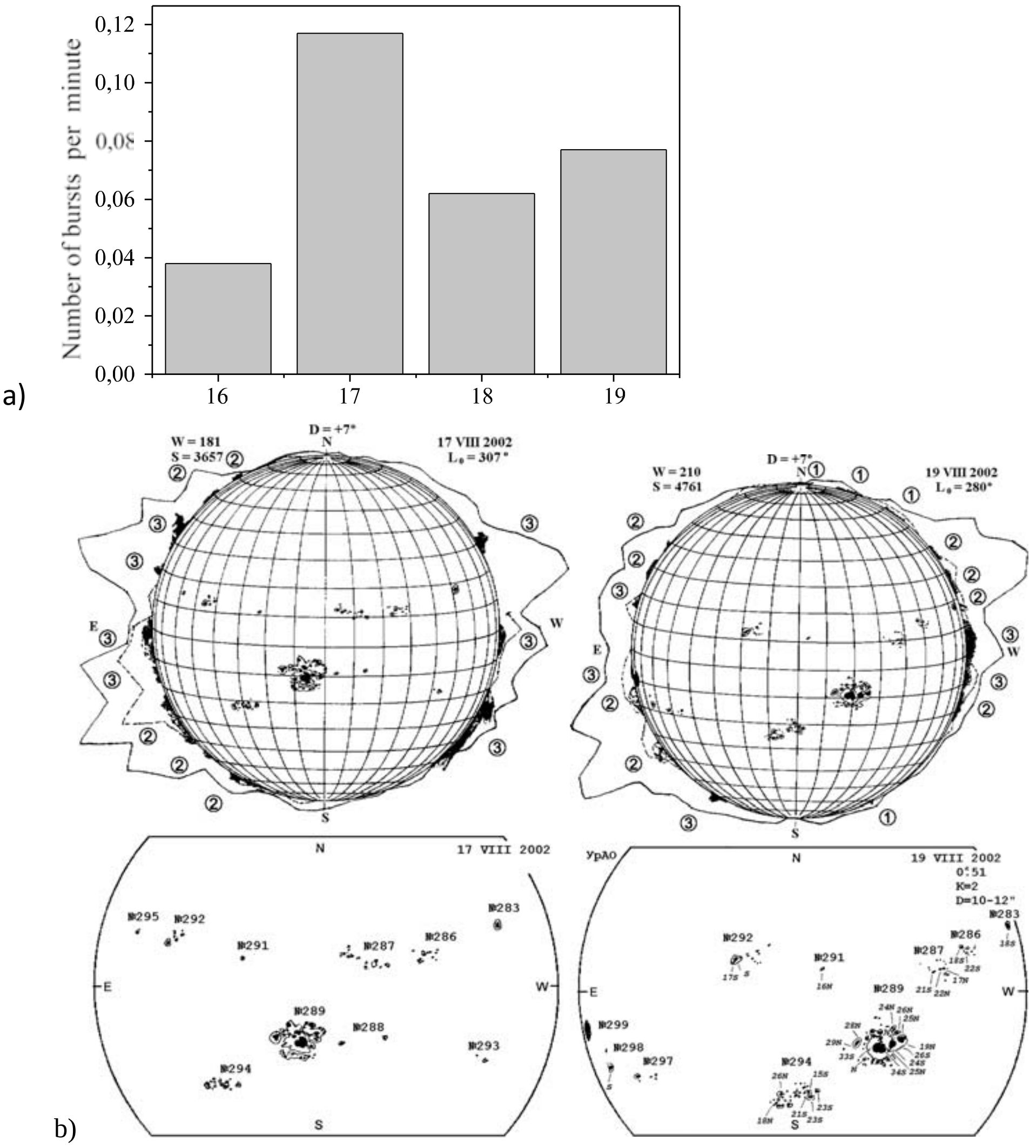

**Figure 5.** (a) The occurrence rate of Type III-like bursts in the presence of two active regions on the solar disk. (b) Two active regions intersecting the central meridian on 17 and 19 August 2002 (http://www.gao.spb.ru/).

The harmonic radio emission is generated by the process $l + l \rightarrow t$, in which two Langmuir waves $l$ coalesce into the electromagnetic wave $t$. From conservation of energy one can get $\omega_{l1} + \omega_{l2} = \omega_t$ in this process (Ginzburg and Zhelezniakov, 1958; Suzuki and Dulk, 1985). The wave number and the group velocity of the generated electromagnetic wave are given by

$$k_t = \sqrt{3}\frac{\omega_{pe}}{c} \tag{10}$$

and

$$v_{gr} = \frac{\sqrt{3}c}{2} \ , \tag{11}$$

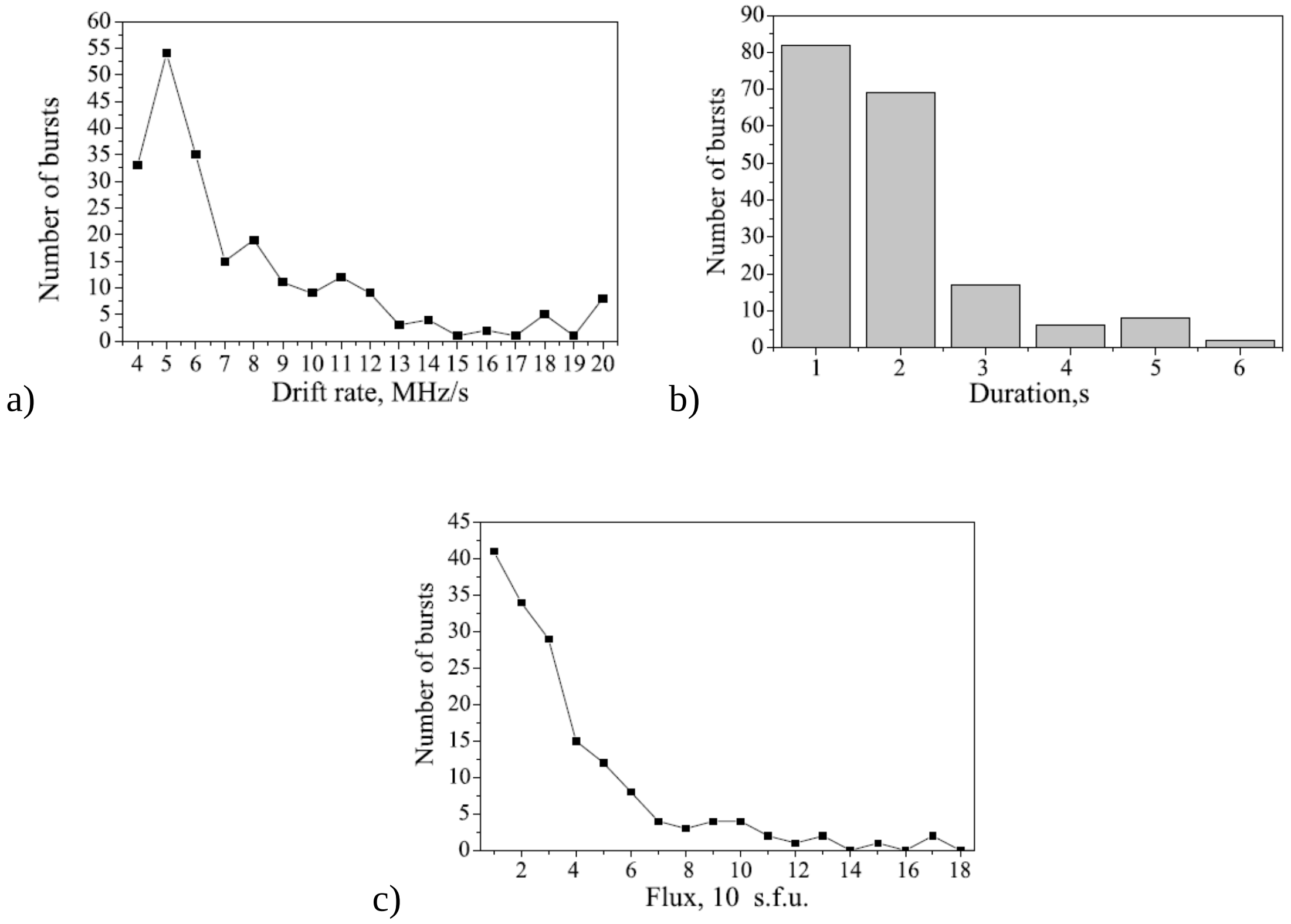

**Figure 6.** The relations between the occurrence rate of Type III-like bursts and (a) drift rates, (b) duration, and (c) radio flux for the storm of 24 – 30 July 2002.

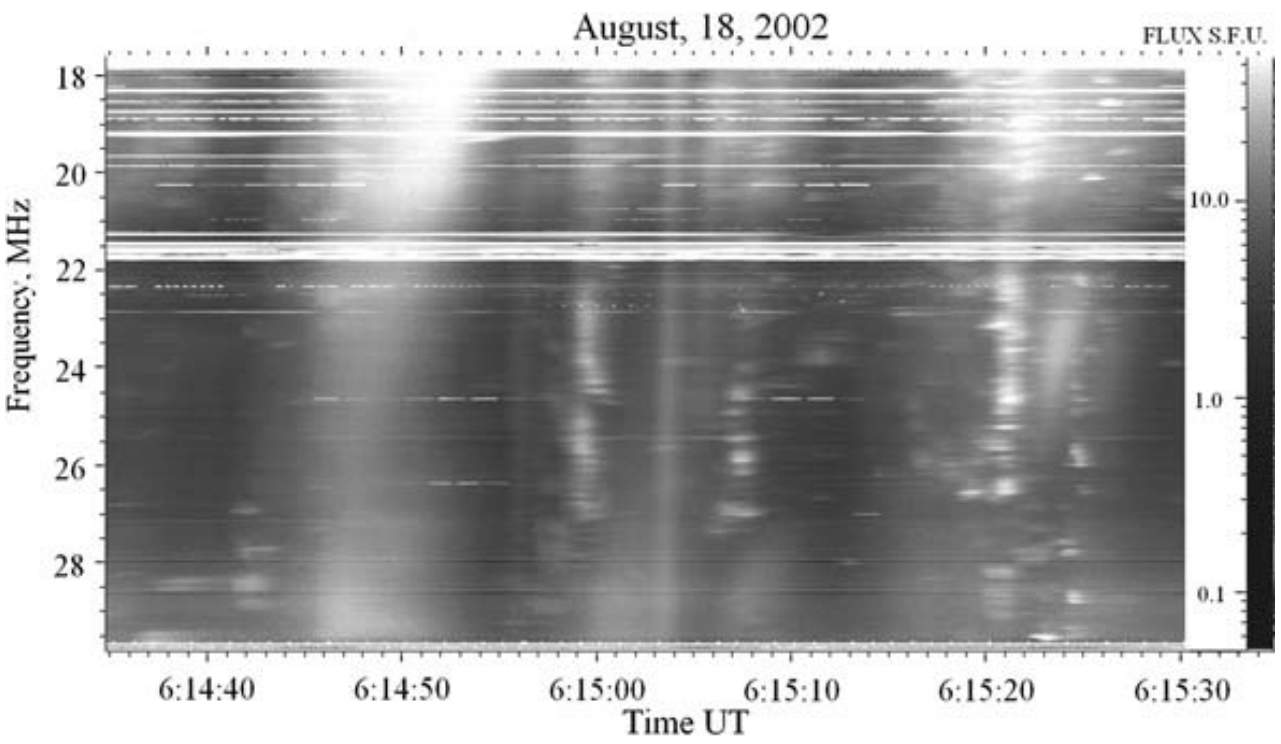

**Figure 7.** The Type III-like burst (06:15:04 UT) with maximum drift rate.

respectively. Using Equations (4) and (11) we derive the critical velocity in this case as

$$v_s{*} = \frac{\sqrt{3}c}{2\cos\alpha} = \frac{2.6 10^{10}}{\cos\alpha} cms^{-1} \tag{12}$$

Therefore, if electrons with velocity $v_s{*} = 510^9\ \mathrm{cm}\ s^{-1}$ travel toward the observer ($\alpha = 0$) then according to Equation (9) $\gamma$ is large and fast Type III bursts can be observed as the fundamental radio emission. In this case the active region is situated near the central meridian. At bigger angles $\alpha$ (when the active region is situated far from the central merid- ian) the velocity $v^*$ becomes larger. Since the number of electrons accelerated in an active region with high velocities is smaller than that with small velocities it is reasonable to suggest that there will be fewer Type III-like bursts when an active region is at some distance from the central meridian. To observe the harmonic radio emission of Type III-like bursts the electrons with higher velocities ($2.6 \times 10^{10}$ cm $\mathrm{s}^{-1}$) are needed. So we see that the presence of fast electrons with subrelativistic velocities is sufficient to understand the phenomenon of Type III-like bursts.

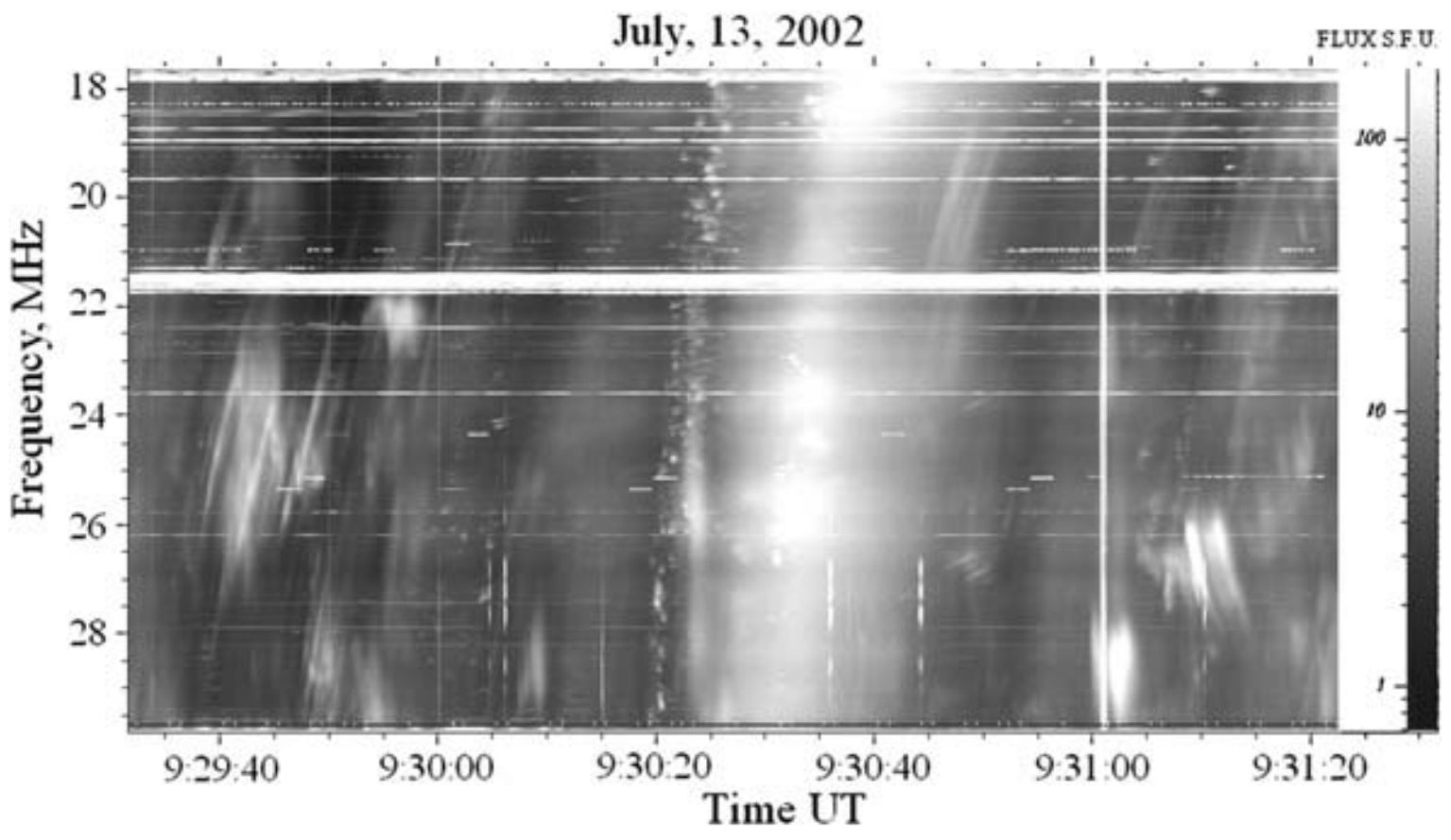

**Figure 8.** Type III-like burst (09:30:24 UT) with positive drift rate observed on 13 July 2002.

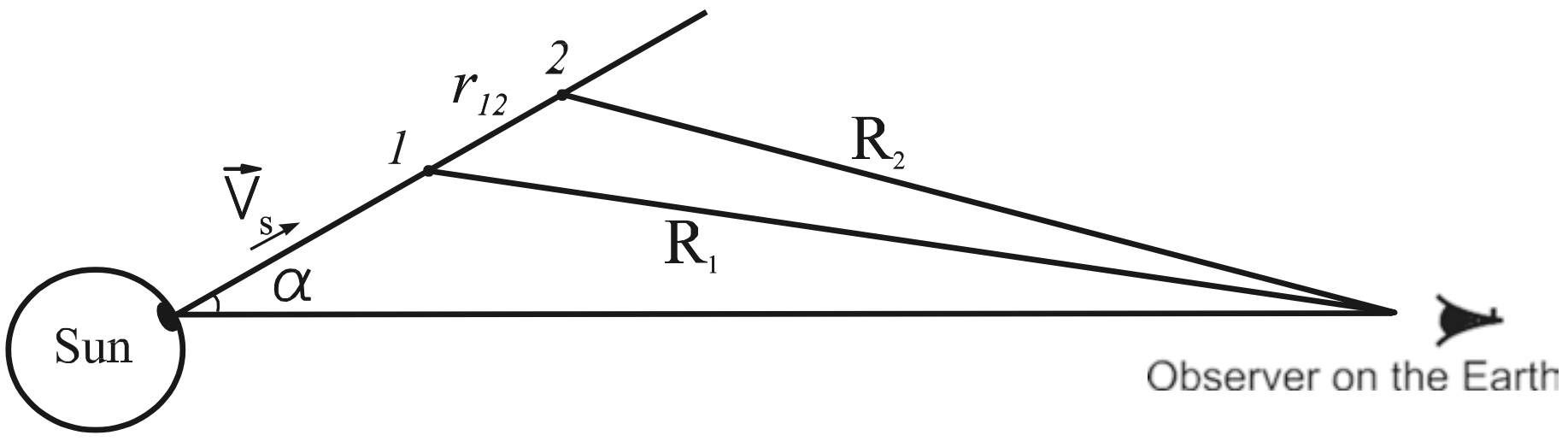

**Figure 9.** Schematic diagram showing the mechanism of the frequency drift. The two waves at frequencies $f_1$ and $f_2$ are generated by Type III electrons at two different spatial points with distance between them $r_{12}$.

All these reasons are correct for the case of the generation of monochromatic Langmuir waves with wave number $k_l = \omega_{pe}/v_s$. But in the real situation a broad spectrum of Langmuir waves is excited. So these results should be reconsidered. It is also very important to take into account the size of the region from which the radio emission emerges, because the electromagnetic waves radiated at the same frequency, but in different places in the corona, pass different propagation paths (*e.g.*, the decameter radio emission occurring in the corona comes from a volume with linear sizes comparable to the solar radius) and this influences the resultant drift rate of the burst. So the values of $v^*$ should be reconsidered in the case of a non-point-like source of radio emission and a real spectrum of Langmuir waves. Such a problem is more adequate to observations and can be solved numerically, but this is beyond the frame of the present paper. Here we have only shown one of the possibilities to understand the phenomenon of Type III-like bursts.

## 4. Conclusions

In this paper we have presented the first results of observations of decameter Type III- like bursts with frequency drift rates substantially larger than those for the standard decameter Type III bursts. Although the drift rates for the standard Type III bursts are about 2 – 4 MHz $s^{-1}$, there are many Type III-like bursts with drift rates of 10 – 30 MHz $s^{-1}$ and even larger. Statistical analysis has shown that the majority of Type III-like bursts have a duration of 1 – 2 s, which is a quarter of that for the usual decameter Type III bursts. Such a ratio is also characteristic for meter – decimeter Type III-like bursts. For high-frequency bursts we found a strong dependence of Type III-like bursts on the location of active regions associated with them. Analyzing five Type III storm events (in different years and months), we found that the maximum number was observed when the active region was situated near the central meridian. Earlier this feature for high-frequency Type III-like bursts was considered to be connected to the propagation effects of electromagnetic waves in the solar corona (Elgaroy, 1980). Not only do we consider these effects to be responsible for this feature but we find that they can also directly clarify the phenomenon of Type III-like bursts. By taking into account both the electron velocity and the group velocity of the electromagnetic waves, we were able to explain the large drift rates of Type III-like bursts. We have also shown that the drift rates can be large and can even change sign. In the special case in which the group velocity of the electromagnetic waves and the electron velocity are equal, we would observe radio emission at different frequencies at the same time (*i.e.*, with a drift rate of infinity). This effect is most pronounced in the case of electrons propagating toward the observer, which corresponds to the situation when an active region is near the central meridian. We derived the critical velocities for fundamental and harmonic radio emissions of Type III- like bursts in the case of a point-like emission source and when a monochromatic Langmuir wave with wave number of $k_l = \omega_{pe}/v_s$ is generated. A further analysis of a more realistic situation of a radio emission region with finite size and a broad Langmuir-wave spectrum will be presented in future publications.

**Acknowledgements** The work was partly supported by the INTAS 03-5727 Project and the Commission for International Cooperation of the Austrian Academy of Sciences. The authors are indebted to the anonymous referee for useful comments.